\documentclass[secnumarabic,amssymb, nobibnotes, aps, prd]{revtex4}%
\usepackage{graphicx}
\usepackage{dcolumn}
\usepackage{bm}
\usepackage{amsmath}%
\usepackage{amsfonts}%
\usepackage{amssymb}

\begin{document}

\title{Faltung Formulation of Hadron Halo Event Cascade
at Mt. Chacaltaya}

\author{K H Tsui$^{\dag}$, H M Portella$^{\dag}$, C E Navia$^{\dag}$, H Shigueoka$^{\dag}$,
 and L C S de Oliveira$^{\ddag}$}
\affiliation{$^{\dag}$Instituto de F\'{i}sica - Universidade Federal Fluminense,
Campus da Praia Vermelha, Av. General Milton Tavares de Souza s/n,
Gragoat\'{a}, 24.210-346, Niter\'{o}i, Rio de Janeiro, Brasil.}

\affiliation{$^{\ddag}$ Centro Brasileiro de Pesquisas F\'{i}sicas CBPF/MCT,
Rua Dr. Xavier Sigaud 150,
Botafogo, 22290-180, Rio de Janeiro, Rio de Janeiro, Brazil.} 

\date{\today}

\begin{abstract}

It is shown that the fundamental standard hadron cascade diffusion
 equation in the Mellin transform space is not rigorously correct
 because of the inconsistent double energy integral evaluation
 which generates the function $<\eta^{s}>$ with its associated
 parametizations. To ensure an exact basic working equation,
 the Faltung integral representation is introduced which has the
 elasticity distribution function $u(\eta)$ as the only fundamental
 input function and $<\eta^{s}>$ is just the Mellin transform of
 $u(\eta)$. This Faltung representation eliminates standard
 phenomenological parameters which serve only to mislead the
 physics of cascade.
 The exact flux transform equation is solved by the method of
 characteristics, and the hadron flux in real space is obtained by
 the inverse transform in terms of the simple and essential residues.
 Since the essential residues are given by the singularities
 in the elasticity distribution and particle production transforms
 that appear in the exponentials, these functions should not be
 parametized arbitrarily to avoid introducing non-physical essential
 residues.
 This Faltung formulation is applied to the charged hadron integrated
 energy spectra of halo events detected at Mt. Chacaltaya by the
 Brazil-Japan Collaboration, where the incoming flux at the
 atmospheric boundary is a single energetic nucleon.

\end{abstract}


\maketitle

\newpage
\section{Introduction}

The Brazil-Japan Collaboration detected many cosmic-ray events in the
 energy range $(10^{13} - 10^{17})\,eV$ with emulsion chamber exposed
 at Mt. Chacaltaya. About twenty events were observed in the visible
 energy region $E~\geq~1000\,TeV$. Approximately, half of them
 ~\cite{lattes, amato, chinellato, yamashita} were associated with a
 uniformly darkened wide area on X-ray films. The central part of this
 area was called "halo", and so these events were called "halo events".
 Similar events were also observed in experiments at
 Pamir~\cite{pamir},Fuji~\cite{akashi}, and Kanbala~\cite{ren}.
 Recently, a new experiment~\cite{aoki}, using a hadron calorimeter
 associated with emulsion chambers at Mt. Chacaltaya, reported
 this kind of events as well.
 Thus, the appearance of a strong concentration of energy and particles
 as a halo seemed to be a common feature in this energy region.
 After the 1980s, these super-families were also compared with
 simulations using different primary compositions and models for
 high-energy nuclear interactions~\cite{capedevielle, kalmikov, werner}.
 Nevertheless, these simulations could not describe fully all the
 events with the same inputs on primary composition and nuclear
 collision models.
 Due to their large hadronic number and energy, some authors~\cite{amato2}
 suggested that these events could be explained as centauro-like ones.

Recently, we have developed an analytic method~\cite{tsui}
 that allows us to calculate the hadronic and electromagnetic
 components of cosmic rays in the Earth's atmosphere, and
 generalizes earlier works~\cite{arata, boyadzhyan, bellandi86, ohsawa}
 by including the energy dependence of the mean-free path effect.
 An important issue in the high energy region concerns the behavior
 of the inelasticity, which is defined as the fraction of energy
 given up by the leading hadron in a collision induced by an incident
 hadron on a target nucleon or nucleus. This parameter has been
 exhaustively studied in several papers, but until now it continues
 to be an open question. Several authors have suggested that the
 average inelasticity coefficient is an increasing function of the
 energy~\cite{deus, kaidalov}, whereas others have proposed that
 it is a decreasing one~\cite{fowler, kotaro, wlodarczyk}.
 High energy cosmic rays, which reflect the nuclear interaction
 in the energy region covering $1~to~100~TeV$, are well fitted
 with a constant mean inelasticity 1/2~\cite{augusto, ohsawa2}.
 At higher energies, where these super-families belong, a constant
 value for the inelasticity is no longer valid in order to explain
 experimental data.
 Up to now, all papers treating analytically this subject have used
 simplified inputs on interaction mean-free paths and inelasticities.

Here, we believe much of the difficulties in developing an analytic
 approach of cascade lies on the fact that the widely used standard
 diffusion equation in Mellin transform space is not a mathematically
 rigorous equation, due to the approximations used in the evaluation
 of the double energy integral that brings in the entity $<\eta^{s}>$.
 Because of this fundamental inconsistency at the very beginning,
 it is reasonable to understand that each level of the cascade would
 compound the array of free parameters. 
 In section~\ref{nucdiffusion}, we make use of the Faltung theorem to
 formulate exactly the diffusion equation in Mellin transform space.
 In this exact Faltung representation, $<\eta^{s}>$ is just
 the transform of the elasticity distribution function $u(\eta)$.
 There is no need to model and parametize $<\eta^{s}>$ itself.
 Consequently, elasticity distribution is the primary and only
 input function.
 As a result, we could avoid over parametizing and contaminating
 a model based on an approximated non-rigorous transform equation.
 Method of characteristics~\cite{tsui} is used to solve the exact
 transform equation considering one single incoming nucleon as the
 boundary condition in section~\ref{seccharac}.
 For the nucleon case, the flux in real space is evaluated by the
 $s=-1$ essential residue in section~\ref{secres}. Our solutions
 are presented in the usual modified Bessel functions of order 1.
 For the pion case, the Faltung representation of the transformed
 equation is treated in section~\ref{secpw}, and the $s=s_{0}$ simple
 and $s=0$ and $s=-1$ essential residues are evaluated in
 section~\ref{secpres}.
 In section~\ref{secconcl}, we compare these fluxes with the halo
 event data of P06, Ursa Maior, Andromeda and Mini-Andromeda III
 ~\cite{amato, chinellato, yamashita} with discussions and conclusions.

\newpage
\section{Faltung Formulation of Nucleon Diffusion}\label{nucdiffusion}

From considerations of different fundamental physical processes,
 the number density flux energy distribution function of nucleons
 $N(E,t)$ at energy $E$ and atmospheric depth $t$ is described by

\begin{eqnarray}
\frac{\partial N(E,t)} {\partial t}
& = & -\frac{N(E,t)} {\lambda(E)}
 +\int_{0}^1 \int_{E}^{\infty} u(\eta) \delta(E-\eta E^{\prime})
 \frac{N(E^{\prime},t)}{\lambda(E^{\prime})}
 dE^{\prime}d\eta \nonumber\\
& = & -\frac{N(E,t)} {\lambda(E)}
 +\int_{E}^{\infty} u(E/E^{\prime})
 \frac{N(E^{\prime},t)}{\lambda(E^{\prime})}
 \frac{1}{E^{\prime}}dE^{\prime} \label{eqno1}
\end{eqnarray}

\noindent where $\lambda(E)$ is the energy dependent mean-free
 path, $\eta=E/E^{\prime}<1$ is the elasticity, $u(E/E^{\prime})$
 is the elasticity distribution. Modelling the mean-free path
 by a power index $\beta$ \cite{portella88},

\begin{equation}
\lambda(E) = \lambda_N (\frac{E}{B})^{-\beta},\label{eqno2}
\end{equation}

\noindent the second term on the right side of Eq.~(\ref{eqno1})
can be converted from a $E^{\prime}$ integral to an $\eta$
integral that reads

\begin{equation}
\frac{\partial N(E,t)}{\partial t}
= -\frac{1}{\lambda_{N}}(\frac{E}{B})^{\beta} N(E,t)  +\frac{1}{\lambda_{N}}(\frac{E}{B})^{\beta}
 \int_{0}^1 (\frac{1}{\eta})^{\beta+1} u(\eta)
 N(\frac{E}{\eta},t)d\eta
\label{eqno3}
\end{equation}

\noindent where $B$ is the normalization energy of the mean-free path.
 We note that the energy dependence in $\lambda(E)$ makes the mean
 free path decrease as energy increases, $(E/B)>1$. The power index
 $\beta>0$ depends on the reference energy $B$.
 We observe that the nucleon cascade equation, Eq.~(\ref{eqno1}),
 has twocompeting terms on the right side. The first term is the
 diffusion term that drains the flux $N(E,t)\,dE$ at $E$ to lower
 energies $E^{\prime}$. The second term is the attenuation term
 that fills the flux at $E$ by higher energies $E^{\prime}$.
 Since the mean-free path scaled by Eq.~(\ref{eqno2}) vanishes as
 $E/B$ goes to infinity with $\beta$, the first term would dominate
 the equation and the spatial gradient of the flux would be very
 negative at high energies.

Instead of solving Eq.~(\ref{eqno3}) in real space
 \cite{boyadzhyan,portella01}, we use the integral transform approach.
 Before doing the $\eta$ integral, we take the Mellin transform
 defined by

\begin{eqnarray} 
 \tilde{N}(s,t) = \int_{0}^{\infty}
(\frac{E}{A})^{s}  N(\frac{E}{A},t)\,d(\frac{E}{A})\label{eqno4a} \\
 N(\frac{E}{A},t) = \frac{1}{2\pi i} \int
(\frac{E}{A})^{-(s+1)}\tilde{N}  (s,t)\, ds \label{eqno4b}
\end{eqnarray}

\noindent where the energy $E$ is normalized to some reference
 energy $A$, so that the transform does not carry dimension of
 energy to power $s$. Now, Eq.~(\ref{eqno3}) in the transform
 space reads

\begin{eqnarray}
\frac{\partial \tilde{N}(s,t)}{\partial t}
 = -\frac{1}{\lambda_{N}}( \frac{A}{B})^{\beta} \tilde{N}(s+\beta,t)
\nonumber \\
 +  \frac{1}{\lambda_{N}}(\frac{A}{B})^{\beta}
 \int_{0}^{\infty} <\eta^{s}>
 (\frac{E/A}{\eta})^{s+\beta} N(\frac{E/A}{\eta},t)
 d(\frac{E/A}{\eta} ) \label{eqno5}
\end{eqnarray}

\noindent To reach this equation, we have done the $\eta$ integral
 locally to get $<\eta^{s}>$ while ignoring the $\eta$ dependent
 terms in the $E$ integral. With $K$ as the normalization energy,
 the following model of parametization is often used \cite{kotaro}

\begin{eqnarray}
<\eta^{s}>   =  \int_{0}^{1}u(\eta)\eta^{s}d\eta 
 = \frac{1}{(1+\delta s)}
 (\frac{E^{\prime}}{K})^{\kappa s} \nonumber \\
  = \frac{1}{(s-s_{0})}\frac{1}{\delta}
 (\frac{A}{K} )^{\kappa s}(\frac{E^{\prime}}{A})^{\kappa s}
  = a(s)(\frac{A}{K} )^{\kappa s}(\frac{E^{\prime}}{A})^{\kappa s}
 \label{eqno6a}
 \end{eqnarray}

\noindent where $s_{0}=-1/\delta$. For a uniform elasticity distribution,
we have $ \kappa=0$, $\delta=1$, and $s_{0}=-1$. In particular, taking
$s=1$ gives the average elasticity

\begin{equation}
<\eta> =
\frac{1}{(1+\delta)}(\frac{E^{\prime}}{K})^{\kappa} \label{eqno6b}
\end{equation}

\noindent Since $<\eta^{s}>$ is parametized to an $E^{\prime}$
 dependence, the $E$ integral is now converted back to an $E^{\prime}$
 integral with the help of $\eta$ to complete the second term.
 The equation of the flux transform then becomes

\begin{eqnarray}
(\frac{A}{B})^{-\beta}\frac{\partial \tilde{N}(s,t)}{\partial t}
& = &  -\frac{1}{\lambda_{N}}\tilde{N}(s+\beta,t)\nonumber\\
 &  & +\frac{1}{\lambda_{N}}(\frac{A}{K})^{\kappa s} a(s)
\tilde{N}(s+\beta+\kappa s,t) \label{eqno7}
\end{eqnarray}

\noindent We notice that, should the energy $E$ in the Mellin
 transform be not normalized to some reference energy $A$,
 then $\tilde{N}(s,t)$, $\tilde{N}(s+\beta,t)$ and
 $\tilde{N}(s+\beta+\kappa s,t)$ would have different dimensions
 in energy. Here, in Eq.~(\ref{eqno5}), they have the same dimension
 of $N(E,0)$.
 We remark that the second term on the right side of Eq.~(\ref{eqno5})
 is obtained by isolating the $<\eta>$ integral to get $<\eta^{s}>$
 and leaving other $\eta$ dependent terms to be converted to the
 $E^{\prime}$ integral.
 Such mathematically unwaranted procedure makes
 Eq.~(\ref{eqno5}) not an exact representation of Eq.~(\ref{eqno1}).
 Consequently, any parametization on the elasticity in the approximated
 flux transform equation could only contaminate the physics of
 the nucleon flux in real space.
 
To avoid this serious mathematical inconsistency,
 we reconsider Eq.~(\ref{eqno1}) and make use of the Faltung integral.
 By defining the elasticity distribution $u(E/E^{\prime})=0$ for
 $E/E^{\prime}>1$, the lower limit of the integral can be extended
 to zero, and Eq.~(\ref{eqno1}) reads

\begin{eqnarray}
\frac{\partial N(E/A,t)} {\partial t}
& = & -\frac{N(E/A,t)} {\lambda(E)}
+\int_{0}^{\infty} u(E/E^{\prime})
 \frac{N(E^{\prime}/A,t)}{\lambda(E^{\prime})}
 \frac{1}{E^{\prime}}dE^{\prime} \nonumber\\
& = & -w(E,t)
+\int_{0}^{\infty} u(E/E^{\prime}) w(E^{\prime},t)
\frac{1}{E^{\prime}}dE^{\prime} \nonumber\\
& = & -w(E,t) + u \otimes w(E,t)\label{eqno7a}
\end{eqnarray}

\noindent The last equality is reached by using the Faltung integral
of the Mellin transform. Taking the Mellin transform leads to

\begin{eqnarray}
\frac{\partial \tilde{N}(s,t)}{\partial t}
& = & -\tilde{w}(s,t)
 + \tilde{u}(s)\tilde{w}(s,t) \nonumber\\
& = & -[1-\tilde{u}(s)]\tilde{w}(s,t) 
 = -\frac{1}{\lambda_{N}}(\frac{A}{B})^{\beta}
 [1-\tilde{u}(s)]\tilde{N}(s+\beta,t)\label{eqno7b}
\end{eqnarray}

\begin{equation}
\tilde{u}(s)
 = \int_{0}^{\infty}(\frac{E}{A})^{s} u(\frac{E}{A})
d(\frac{E}{A})  =  <\eta^{s}> \label{eqno7c}
\end{equation}

\noindent By changing the normalization energy $A$ in Eq.~(\ref{eqno7c})
 to $E^{\prime}$, and recalling $u(E/E^{\prime})=0$ for
 $E/E^{\prime}>1$, we see that $<\eta^{s}>$ is just the
 Mellin transform of the elasticity distribution function.
 It is completely determined once elasticity distribution is specified.

In this Faltung representation, elasticity distribution $u(\eta)$
 is the fundamental parameter. There is no need to specify $<\eta^{s}>$
 together with those unnecessary parameters associated to it.
 The necessity of parametizing $<\eta^{s}>$ is actually derived from
 the approximated nature of Eq.~(\ref{eqno5}).
 This could over parametize the model based on an
 approximated equation, which could generate inconsistencies among
 other parameters. An exact representation of Eq.~(\ref{eqno1}) by
 Eq.~(\ref{eqno7b}) through Faltung theorem sets a firm base upon
 which additional features of one-dimensional cascade could be
 analyzed with confidence.
 This Faltung representation apparently requires $u(E/E^{\prime})$
 as a function of $E/E^{\prime}$ only. However, this is not true.
 With $\eta$ and $E^{\prime}$ dependences in a separable form,
 we could transfer the $E^{\prime}$ dependence of
 $u(\eta,E^{\prime})$ to $w(E^{\prime},t)$. By defining a new
 $w(E^{\prime},t)$, the Faltung formulation prevails.

\newpage
\section{Method of Characteristics}\label{seccharac}

With the Faltung formulation, we proceed to solve Eq.~(\ref{eqno7b}).
 Some researchers solve equation of this kind formally by
 operators \cite{bellandi90}. Since $\beta$ is much less than $s$,
 we choose to make a Taylor expansion of $\tilde{N}(s+\beta,t)$
 to get a first order differential equation
\begin{equation}
\lambda_{N}(\frac{A}{B})^{-\beta}
\frac{\partial \tilde{N}(s,t)}{\partial t}
+[1-\tilde{u}(s)]\beta
\frac{\partial \tilde{N}(s,t)}{\partial s}\,
=\,-[1-\tilde{u}(s)] \tilde{N}(s,t) \label{eqno8}
\end{equation}

\noindent This partial differential equation is equivalent to the following
set of ordinary differential equations which describes the trajectory of
the coordinate point $(s,\, t,\,\tilde{N})$ in the functional space
parameterized to $\xi \, $ \cite{courant,tsui92,tsui93}

\begin{equation}
(\frac{A}{B})^{\beta}\frac{d\,t}{\lambda_{N}} =
\frac{d \, s}{[1-\tilde{u}(s)]\beta}  =
-\frac{d\tilde{N}}{[1-\tilde{u}(s)]\tilde{N}} = d\xi. \label{eqno9}
\end{equation}

\noindent This method of characteristics to solve first order partial
differential equations was used in superradiant free electron
lasers \cite{tsui92,tsui93}. Solving  for the equality between $dt$
and $ds$,

\begin{equation}
(\frac{A}{B})^{\beta}\frac{d \, t}{\lambda_{N}} =
\frac{d\,s}{[1-\tilde{u}(s)]\beta} \label{eqno10}
\end{equation}

\noindent we get a trajectory between the variables $t$ and $s$ through
the parameter $\xi$, $t=t(s,\beta,\kappa)$, which is the characteristics
of the partial differential equation, Eq.~(\ref{eqno8}).

To get the transform of the flux, we could solve the equality of
$d\tilde{N}$ with $d\xi$, or with $ds$, or with $dt$. Since the boundary
condition of $\tilde{N}$ is given in terms of $s$ at $t=0 $, we choose
to solve with $dt$

\begin{equation}
(\frac{A}{B})^{\beta}\frac{d\,t}{\lambda_{N}}
 =  -\frac{d\tilde{N}}{[1-\tilde{u}(s)]\tilde{N}}\nonumber\\
\end{equation}

\begin{equation}
 \tilde{N}(s,t)
 = \tilde{N}(s,0)\,e^{-(A/B)^{\beta}
 [1-\tilde{u}(s)]t/\lambda_{N}} \label{eqno11}
\end{equation}

\noindent For the boundary condition, we use a single incident nucleon
 with energy $E_{0}$ as follows plus its associated Mellin transform

\begin{equation}
N(\frac{E}{A},0) = \delta (\frac{E}{A} - \frac{E_{0}}{A}), \label{eqno12a}
\end{equation}

\begin{equation}
\tilde{N}(s,0)\,= (\frac{A}{E_{0}})^{-s}. \label{eqno12b}
\end{equation}

\noindent The inverse transform of Eq.~(\ref{eqno11}), therefore,
 gives the nucleon flux in real space

\begin{equation}
N(\frac{E}{A},t) = (\frac{E_{0}}{A})^{-1} \frac{1}{2\pi i}
 \int (\frac{E}{E_{0}})^{-(s+1)}
 e^{-(A/B)^{\beta}[1-\tilde{u}(s)]t/\lambda_{N}} ds. \label{eqno14}
\end{equation}

\noindent For the elasticity distribution, which is our fundamental
 parameter, let us consider a linear profile

\begin{equation}
u(\eta) = 1. \label{eqno15}
\end{equation}

\noindent By taking the Mellin transform, we get
  
\begin{equation}
\tilde{u}(s) = \frac{1}{s+1}.  \label{eqno16}
\end{equation}

\noindent We note that the parameter $\beta$ which measures the mean
 free path effects appears in the $(A/B)^{\beta}$ factor and in the
 characteristics of the partial differential equation.

\newpage
\section{Calculus of Residues}\label{secres}

To get the nucleon flux in real space, we need to evaluate the
 residues in Eq.~(\ref{eqno14}). An inspection of the equation
 tells that there are no simple poles. However, in the exponent,
 there is a pole at $s=-1$ in $\tilde{u}(s)$
 
\begin{equation}
\tilde{u}(s) = \frac{1}{s+1}  \label{eqno18}
\end{equation}

Since it appears in the exponent, this is an essential singularity.
 The only contribution to the nucleon flux, therefore, comes from
 the $s=s_{0}=-1$ essential singularity. To consider this essential
 residue, we expand the exponential function in power series to obtain

\begin{eqnarray}
2\pi i Res(s_{0})& = & (\frac{E_{0}}{A})^{-1}
\,  \{(\frac{E}{E_{0}})^{-1}\,e^{-(A/B)^{\beta}t/\lambda_{N}}
\nonumber \\
&  & \sum^{\infty}_{n=0} \frac{1}{n!}
 [(\frac{A}{B})^{\beta}\frac{t}{\lambda_{N}}]^{n}
 \int (\frac{E_{0}}{E})^{s}
 \frac{1}{(s-s_{0})^{n}}\,ds\}. \label{eqno19}
\end{eqnarray}
 
\noindent We define the function acting on the $(s - s_{0})$ powers
 inside the integral by $G(s) = (E_{0}/E)^{s} = g^{s}$.
 Since $G(s)$ is analytic in the neighborhood of $s = s_{0}$,
 we expand it in a Laurent series about $s = s_{0}$ so that
 $Res(s_{0})$ becomes

\begin{eqnarray}
2\pi i Res(s_{0})& = & (\frac{E}{A})^{-1}
\,  \{\,e^{-(A/B)^{\beta}t/\lambda_{N}}
\nonumber \\
&  & \sum^{\infty}_{n=0} \sum^{\infty}_{m=0}
 \frac{1}{n!} \frac{1}{m!}
 [(\frac{A}{B})^{\beta}\frac{t}{\lambda_{N}}]^{n}
 G^{(m)}(s_{0})
 \int \frac{1}{(s-s_{0})^{n-m}}\,ds\}. \label{eqno20}
\end{eqnarray}

\noindent By taking $(n-m)=+1$ terms, we pick up the contributions
to the essential residue so that

\begin{eqnarray}
N(\frac{E}{A},t)& = & (\frac{E}{A})^{-1}
 \,  (\frac{E}{E_{0}})^{-s_{0}}
 \,e^{-(A/B)^{\beta}t/\lambda_{N}}\nonumber\\
 &   & [(\frac{A}{B})^{\beta}\frac{t}{\lambda_{N}}]
 \sum^{\infty}_{n=1} \frac{1}{n!} \frac{1}{(n-1)!}
 (\frac{1}{4}Z^{2})^{n-1}. \label{eqno21}
\end{eqnarray}

\noindent Here $Z^2=4(A/B)^{\beta}(t/\lambda_{N})(\ln g)$.
 The infinite series can be rewritten in terms of the modified
 Bessel function of order one. Thus, the result is

\begin{eqnarray}
N(\frac{E}{A},t)& = & (\frac{E}{A})^{-1}
 \,  (\frac{E}{E_{0}})^{-s_{0}}
 \,e^{-(A/B)^{\beta}t/\lambda_{N}}\nonumber\\
 &   & [(\frac{A}{B})^{\beta}\frac{t}{\lambda_{N}}]
 \frac{2}{Z} I_{1}(Z). \label{eqno27}
\end{eqnarray}

We recall that the series that gives rise to the modified Bessel
 function of order one is a semi-divergent series. It diverges
 up to some $n\,th$ term due to the factors like $(t/\lambda_{N})$
 and $(\ln g)$. Afterwards, it begins to converge due to the
 factorials of $n$. For the single incident nucleon case,
 the fraction $(E_{0}/E)>1$ so that $g$ is larger than unity and
 $(\ln g)$ is positive. Consequently, $Z^{2}(n)$ is positive
 which leads to the modified Bessel function solution.
 The essential residue here plays a very important role, because it
 represents the flux at a given atmospheric depth $t/\lambda_{N}$.
 We remark that this case of single incident nucleon had been solved
 in real space in terms of probability distributions under the
 assumption of constant mean-free path and uniform elasticity
 \cite{arata,boyadzhyan}. It was also solved with Mellin transform
 by residues under the same assumption \cite{ohsawa}. In both
 approaches, the results are in terms of the modified Bessel
 functions $I_{1}(Z)$. To obtain the integrated flux, we use

\begin{equation}
\textbf{N}(\frac{E}{A},t) 
 = \int_{E/A}^{E_{0}/A}N(\frac{E}{A},t)\,d(\frac{E}{A}).\nonumber
\end{equation}

\newpage
\section{Pion Diffusion}\label{secpw}

For the pions, the number density flux energy distribution function
 $\Pi(E,t)$ at energy $E$ and atmospheric depth $t$ is given by

\begin{eqnarray}
\frac{\partial \Pi(E,t)} {\partial t}
& = & -\frac{\Pi(E,t)} {\lambda^{'}(E)}
 +(1-b)\int_{0}^{\infty} u(E/E^{\prime})
 \frac{\Pi(E^{\prime},t)}{\lambda^{'}(E^{\prime})}
 \frac{1}{E^{\prime}}dE^{\prime}
 \nonumber \\
& + & \int_{0}^{\infty} v_{\pi}(E/E^{\prime})
 \frac{\Pi(E^{\prime},t)}{\lambda^{'}(E^{\prime})}
 \frac{1}{E^{\prime}}dE^{\prime}
 +\int_{0}^{\infty} v_{n}(E/E^{\prime})
 \frac{N(E^{\prime},t)}{\lambda(E^{\prime})}
 \frac{1}{E^{\prime}}dE^{\prime} \label{eqno28}
\end{eqnarray}

\noindent Here $b=1/3$ is the charge exchange probability of the
 incident pion, $v(E/E^{\prime})/E^{\prime}=\phi(E^{\prime},E)$,
 and $\phi(E^{\prime},E)dE$ is the number of produced particles
 at energy $E$ in the interval $dE$ due to the incident particle
 at energy $E^{\prime}$. Furthermore, we have again defined
 $u(E/E^{\prime})=0$ and $v(E/E^{\prime})=0$ for $\eta=E/E^{\prime}>1$.
 We also have assummed that $v(E^{\prime},E)=v(E/E^{\prime})$.
 This allows us to use the same Faltung formulation to write
  
\begin{equation}
\frac{\partial \Pi(E,t)} {\partial t}
 =  -w_{\pi}(E,t)
 +(1-b)u \otimes w_{\pi}(E,t)+ v_{\pi} \otimes w_{\pi}(E,t)
 + v_{n} \otimes w(E,t) \label{eqno29}
\end{equation}

\noindent The Mellin transform of the pion equation reads

\begin{eqnarray}
\frac{\partial \tilde{\Pi}(s,t)}{\partial t}
& = & -\tilde{w}_{\pi}(s,t)
 + (1-b)\tilde{u}(s)\tilde{w}_{\pi}(s,t) 
 + \tilde{v}_{\pi}(s)\tilde{w}_{\pi}(s,t)
 + \tilde{v}_{n}(s)\tilde{w}(s,t) \nonumber\\
& = & -[1-(1-b)\tilde{u}(s)-\tilde{v}_{\pi}(s)]\tilde{w}(s,t)
 + \tilde{v}_{n}(s)\tilde{w}(s,t) \label{eqno30}
\end{eqnarray}

\noindent Again, we model the mean-free path of pions by the same
 power index $\beta$

\begin{equation}
\lambda^{'}(E) = \lambda_{\pi} (\frac{E}{B})^{-\beta}\label{eqno31}
\end{equation}

\noindent As for the pion and nucleon production by incident pion,
 there is the Feymann scaling which reads

\begin{equation} 
 \phi(E^{\prime},E)
 =  D[1-(\frac{E}{E^{\prime}})]^d\frac{1}{E} \label{eqno32} 
\end{equation}

\noindent where $D=(d+1)/3$, $d=4$. This gives
 $\phi(E^{\prime},E)E^{\prime}=v(E/E^{\prime})$ which allows the
 application of the Faltung theorem.

Using the method of characteristics with the boundary condition
 $\tilde{\Pi}(s,0)=0$, the pion flux transform is 

\begin{eqnarray}
\tilde{\Pi}(s,t)
 & = & \frac{Q(s)/\lambda_{N}}
 {P(s)/\lambda_{\pi}-\mu(s+\beta)/\lambda_{N}}
 \tilde{N}(s+\beta,t) \nonumber\\
 & \times &
 \{1-e^{+(A/B)^{\beta}\mu(s+\beta)t/\lambda_{N}}
 e^{-(A/B)^{\beta}P(s)t/\lambda_{\pi}}\} \label{eqno33}
\end{eqnarray}

\begin{equation}
\mu(s)=[1-\tilde{u}(s)] \label{eqno34}
\end{equation}

\begin{equation}
P(s)=[1-(1-b)\tilde{u}(s)-Q(s)] \label{eqno35}
\end{equation}

\begin{equation} 
Q(s)=\tilde{v}(s) \label{eqno36}
\end{equation}

\noindent This pion transform is expressed in terms of the nucleon
 transform which acts as a source term. Here, we note that $\beta$
 appears explicitly in $\mu(s+\beta)$ and $\tilde{N}(s+\beta,t)$.
 The mean free path effects are explicit in the pion flux.
 Doing the inverse transform gives the pion flux in real space

\begin{eqnarray}
\Pi(\frac{E}{A},t)
 & = & \frac{1}{2\pi i}(\frac{E_{0}}{A})^{-1}
 \int(\frac{E}{E_{0}})^{-(s+1)}
 \frac{Q(s)/\lambda_{N}}
 {-P(s)/\lambda_{\pi}+\mu(s+\beta)/\lambda_{N}} \nonumber\\
 & \times & \{-e^{-(A/B)^{\beta}\mu(s+\beta)t/\lambda_{N}}
 +e^{-(A/B)^{\beta}P(s)t/\lambda_{\pi}}\} 
 \, ds  \nonumber\\
 & = & \frac{1}{2\pi i}(\frac{E_{0}}{A})^{-1}
 \int(\frac{E}{E_{0}})^{-(s+1)}
 Z(s,\beta) \nonumber\\
 & \times & \{-e^{-(A/B)^{\beta}\mu(s+\beta)t/\lambda_{N}}
 +e^{-(A/B)^{\beta}P(s)t/\lambda_{\pi}}\} 
 \, ds \label{eqno37}
\end{eqnarray}

\noindent As has been discussed in an earlier publication \cite{tsui},
 the pion flux is given by the residues of the simple singularities
 in $Z(s,\beta)$ and the residues of the essential singularities in
 the exponents $\mu(s+\beta)$ and $P(s)$. These essential singularities
 are expressed in an infinite series with terms involving higher and
 higher derivatives. For the nucleon case, there happens to be a closed
 analytic form for the essential residues as in Eq.~(\ref{eqno27}).
 In most cases, we have to evaluate them numerically.

Since the essential residues depend on the singularities of
 $\phi(E^{\prime},E)$ in the exponent, the functional form of
 $\phi(E^{\prime},E)$ acquires a real physical importance,
 beyond simple data fitting.
 For example, should we choose Eq.~(\ref{eqno32}), we would have
 $v(\eta)=\phi(E^{\prime},E)E^{\prime}=(5/3)(1-\eta)^4/\eta$ and

\begin{equation} 
Q(s)=\tilde{v}(s)=D\sum^{4}_{n=0} (-1)^n C^{4}_{n}\frac{1}{(s+n)}
 \label{eqno38}
\end{equation}

\noindent There would be five essential singularities
 s=0,-1,-2,-3,-4, and each singularity would generate
 a pion flux represented by its residue.
 Some of the residues are extraordinarily large and others are
 extraordinarily small. In our opinion, these residues are due to
 singularities that are introduced by data fitting, whose fluxes
 in real space are unphysical. We could also choose

\begin{equation}
v(\eta)\,=\,2.08\frac{1}{\eta}(1-\eta)e^{-5\eta} \nonumber
\end{equation}

\noindent This would generate a Gamma function dependence
 in $\tilde{v}(s)$ which would give an infinite sequence
 of singularities on the left side of the complex $s$ plane.
 Consequently, under the method of residues, arbitrary fitting of
 $\tilde{v}(s)$ could contaminate the flux.

In order to bring out the physics, we consider the energy
 conservation between the average leader pion and the average
 secondary pion energies

\begin{equation} 
<\eta>+<K>\,
 =\,\int [\eta u(\eta)+\eta v(\eta)] d\eta \,=\, 1 \label{eqno40}
\end{equation}

\noindent To obtain the distributions $u(\eta)$ and $v(\eta)$,
 we note that the integrant should satisfy
 
\begin{equation} 
\eta [u(\eta)+ v(\eta)] = 1, 2\eta, 3\eta^2, ... \nonumber
\end{equation}

\noindent Taking the first choice on the right side, a consistent
 pair of distributions is
 
\begin{equation} 
u(\eta) = (1 - a\eta)  \label{eqno42a}
\end{equation}

\begin{equation}
v(\eta) = \frac{1}{\eta}(1 - \eta + a\eta^2)  \label{eqno42b}
\end{equation}

\noindent with elasticity and inelasticity given by respectively

\begin{equation} 
<\eta>\,=\,\int \eta u(\eta) d\eta \,
 =\,\frac{1}{2}-\frac{a}{3} \label{eqno43a}
\end{equation}

\begin{equation} 
<K>\,=\,\int \eta v(\eta) d\eta \,
 =\,\frac{1}{2}+\frac{a}{3} \label{eqno43b}
\end{equation}

\noindent Taking the Mellin transform of $u(\eta)$ and $v(\eta)$
 gives

\begin{equation} 
\tilde{u}(s)
 =[\frac{1}{(s+1)} - \frac{a}{(s+2)}]  \label{eqno44a}
\end{equation}

\begin{equation} 
Q(s)=\tilde{v}(s)
 =[\frac{1}{(s+0)} - \frac{1}{(s+1)} + \frac{a}{(s+2)}] \label{eqno44b}
\end{equation}

\noindent This energy conservation consideration identifies
 two singularities of $\tilde{v}(s)$ at s=0,-1 with $a=0$.
 Again, the assumption that $v(E^{\prime},E)=v(E/E^{\prime})$
 appears to be restrictive. The energy dependence of $v$ can be
 introduced in the Faltung representation in the same way as $u$
 which has been discussed earlier.

\newpage
\section{Pion Residues}\label{secpres}

To evaluate the pion flux of Eq.~(\ref{eqno37}), we have to consider
 the simple residues of $Z(s,\beta)$ and the essential rresidues of
 $\mu(s+\beta)$ and $P(s)$ in the exponents. The nature of these
 residues have been discussed before \cite{tsui}.
 For the present case with $\tilde{u}(s)$ and $\tilde{v}(s)$ given
 by Eq.~(\ref{eqno44a}) and Eq.~(\ref{eqno44b}) respectively
 with $a=0$, using $\alpha=\lambda_{N}/\lambda_{\pi}=5/7$, and
 writing $Z(s,\beta)=Z(s)$ by neglecting the mean free path
 effects in this algebric function, the simple poles of $Z(s)$
 with $b=1/3$ are given by a pair of complex conjugate roots

\begin{equation} 
(s_{0}, s_{0}^{*}) = (s_{r}\pm is_{i}) 
 = \frac{1}{12}(5\pm i(335)^{1/2}) \label{eqno46}
\end{equation}

\noindent As for the exponential function, the mean free path
 effects are represented by $\mu(s+\beta)$ in the first exponent
 of Eq.~(\ref{eqno37}). Here, even $\beta$ is retained to evaluate
 the exponential, the effects are still small on a Log-Log plot.
 Evaluating $\mu(s)$ and $P(s)$ at these roots give respectively

\begin{equation}
(\mu_{r}\pm i\mu_{i})
 = 1-\frac{1}{|s_{0}+1|^2}((s_{r}+1)\mp is_{i})\label{eqno47}
\end{equation}

\begin{equation}
(P_{r}\pm iP_{i}) = 1-\frac{1}{|s_{0}|^2}(s_{r}\mp is_{i})
 +\frac{1}{|s_{0}+1|^2}((s_{r}+1)\mp is_{i})\label{eqno48}
\end{equation}

\noindent With these notations, the simple residues are

\begin{eqnarray}
2\pi i(Res(s_{0})+Res(s_{0}^{*})) 
 & = & (\frac{E_{0}}{A})^{-1}(\frac{E}{E_{0}})^{-(s_{r}+1)}
 \frac{1}{(1-\alpha)}\frac{1}{s_{i}} \nonumber \\
 & \times & \{e^{-\mu_{r}(A/B)^{\beta}(t/\lambda_{N})}
 \sin[\mu_{i}(\frac{A}{B})^{\beta}\frac{t}{\lambda_{N}}
 +s_{i}\ln (\frac{E}{E_{0}})]
 \nonumber \\
 & - & e^{-P_{r}(A/B)^{\beta}(t/\lambda_{\pi})}
 \sin[P_{i}(\frac{A}{B})^{\beta}\frac{t}{\lambda_{\pi}}
 +s_{i}\ln (\frac{E}{E_{0}})]\}
 \label{eqno49}
\end{eqnarray}

\noindent These simple residues have a $(E/E_{0})^{-(s_{r}+1)}$
 power dependence on energy, plus a $\ln (E/E_{0})$ energy dependence
 in the argument of the sine function. Because of these dependences,
 this part of the differential pion flux on a Log-Log plot is
 a straight line with $-(s_{r}+1)$ slope plus a slight modulation
 due to the sine function.

As for the essential residues, let us rewrite Eq.~(\ref{eqno37}) as

\begin{eqnarray}
\Pi(\frac{E}{A},t)
 & = & -\frac{1}{2\pi i}(\frac{E_{0}}{A})^{-1}
 e^{-(A/B)^{\beta}(t/\lambda_{N})}
 \int(\frac{E}{E_{0}})^{-(s+1)}Z(s)
 e^{+\tilde{u}(s+\beta)(t/\lambda_{N})} \, ds  \nonumber\\
 &   & +\frac{1}{2\pi i}(\frac{E_{0}}{A})^{-1}
 e^{-(A/B)^{\beta}(t/\lambda_{\pi})}
 \int(\frac{E}{E_{0}})^{-(s+1)}Z(s)
 e^{+[(1-b)\tilde{u}(s)+\tilde{v}(s)](t/\lambda_{\pi})} \, ds  \nonumber
\end{eqnarray}

\noindent Substituting the expressions of $\tilde{u}(s)$ and
 $\tilde{v}(s)$, we get

\begin{eqnarray}
\Pi(\frac{E}{A},t)
 & = & -\frac{1}{2\pi i}(\frac{E_{0}}{A})^{-1}
 e^{-(A/B)^{\beta}(t/\lambda_{N})}
 \int(\frac{E}{E_{0}})^{-(s+1)}Z(s)
 e^{+(A/B)^{\beta}(1/(s+\beta+1))(t/\lambda_{N})} \, ds  \nonumber\\
 &   & +\frac{1}{2\pi i}(\frac{E_{0}}{A})^{-1}
 e^{-(A/B)^{\beta}(t/\lambda_{\pi})}
 \int(\frac{E}{E_{0}})^{-(s+1)}Z(s) \nonumber\\
 &   & \times e^{+(A/B)^{\beta}(+1/s)(t/\lambda_{\pi})}
 e^{+(A/B)^{\beta}(-b/(s+1))(t/\lambda_{\pi})} \, ds  \nonumber\\
 & = & -\frac{1}{2\pi i}(\frac{E_{0}}{A})^{-1}
 e^{-(A/B)^{\beta}(t/\lambda_{N})}
 \int(\frac{E}{E_{0}})^{-(s+1)}Z(s)
 W_{1N}(s,x) \, ds  \nonumber\\
 &   & +\frac{1}{2\pi i}(\frac{E_{0}}{A})^{-1}
 e^{-(A/B)^{\beta}(t/\lambda_{\pi})}
 \int(\frac{E}{E_{0}})^{-(s+1)}Z(s)
 W_{0\pi}(s,x) W_{1\pi}(s,x) \, ds \label{eqno50}
\end{eqnarray}

\noindent The exponent associated to $\lambda_{N}$ in the first term
 has a pole at $(s+\beta+1)=(s+1)=0$. Following the same mathematical
 procedures of nucleon flux of Eq.~(\ref{eqno21}), we have the partial
 flux

\begin{eqnarray}
2\pi i ResN(-1) 
 & = & -(\frac{E}{A})^{-1} e^{-(A/B)^{\beta}(t/\lambda_{N})} \nonumber \\
 &   & \times \{\sum^{\infty}_{m=0} \frac{1}{m!} \frac{1}{(m+1)!}
 [(\frac{A}{B})^{\beta}\frac{t}{\lambda_{N}}]^{m+1}
 G_{0N}^{(m)}(s,x)\} \label{eqno51}
\end{eqnarray}

\begin{equation}
 G_{0N}(s,x) = (\frac{E_{0}}{E})^{s}Z(s) \label{eqno52}
\end{equation}

\noindent The two exponents associated to $\lambda_{\pi}$ in the
 second term have poles at $(s+1)=0$ and $s=0$, and the partial
 fluxes are

\begin{eqnarray}
2\pi i Res\pi(-1)
 & = & +(\frac{E}{A})^{-1} e^{-(A/B)^{\beta}(t/\lambda_{\pi})} \nonumber \\
 &   & \times \{\sum^{\infty}_{m=0} \frac{1}{m!} \frac{1}{(m+1)!}
 [(\frac{A}{B})^{\beta}\frac{t}{\lambda_{\pi}}(-b)]^{m+1}
 G_{0\pi}^{(m)}(s,x)\} \label{eqno53a}
\end{eqnarray}

\begin{eqnarray}
2\pi i Res\pi(0)
  =  +(\frac{E}{A})^{-1} e^{-(A/B)^{\beta}(t/\lambda_{\pi})} \nonumber \\
     \times \{\sum^{\infty}_{m=0} \frac{1}{m!} \frac{1}{(m+1)!}
 [(\frac{A}{B})^{\beta}\frac{t}{\lambda_{\pi}}]^{m+1}
 G_{1\pi}^{(m)}(s,x))\} \label{eqno53b}
\end{eqnarray}

\begin{equation}
 G_{(0,1)\pi}(s,x) = (\frac{E_{0}}{E})^{s}Z(s)W_{(0,1)\pi}(s,x) \label{eqno54}
\end{equation}

\noindent Since the essential residues are derived from $\tilde{u}(s)$
 and $\tilde{v}(s)$, the functions $u$ and $v$ have real physical
 significance. They directly determine the flux in real space.
 For this reason, $u$ and $v$ must not be arbitrarily fitted to data,
 much less in raising the power to avoid generating artificial
 residues. The integrated pion flux is given by

\begin{equation}
\textbf{P}(\frac{E}{A},t)
 = \int_{E/A}^{E_{0}/A} \Pi(\frac{E}{A},t)\,d(\frac{E}{A}).\nonumber
\end{equation}

\newpage
\section{Results and conclusions}\label{secconcl}

Figure 1 shows the integral hadron fluxes of four halo events
 detected at Mt. Chacaltaya by the Brazil-Japan Collaboration.
 In order to make numerical calculations about the integral hadron
 flux with our model and to compare with the detected events,
 we take the mean-free path normalization energy
 $B=1\,TeV$, $\lambda_{N}=80\,g/cm^{2}$ and $\beta = 0.06$
 which are obtained from accelerator and EAS data in the region
 $1\,TeV \leq E_{lab} \leq 1000\,TeV$ \cite{portella88}.
 For the pion mean-free path, we assume that
 $\lambda_{N}/\lambda_{\pi}=5/7$, and that it has the same energy
 dependence like the nucleon case,
 Eq.~(\ref{eqno2}) \cite{portella98,portella04}.
 In the present calculations, we have only one free parameter,
 $A(TeV)$, which is the normalization energy in the Mellin transform,
 Eq.~(\ref{eqno4a}) and Eq.~(\ref{eqno4b}). This parameter is used
 to repreesnt the normalized energy $E/A$. In order to be compatible
 to the horizontal axis of Fig.1, we choose $A=E_{0}=1000\,TeV$
 so that $E/A\leq 1$. With all these parameters and considering
 a uniform elasticity distribution and constant mean free path
 with $\beta=0$, Fig.2 shows the integrated nucleon flux
 labelled by N from Eq.~(\ref{eqno27}),
 the integrated pion flux from nucleon labelled by PN due to
 the $s=-1$ essential residue from Eq.~(\ref{eqno51}),
 the integrated pion flux from pion labelled by PPe due to
 the $s=0$ essential residue from Eq.~(\ref{eqno53b}),
 the integrated pion flux from pion labelled by PPs due to
 the $s=s_{0},s^{*}_{0}$ simple residues from Eq.~(\ref{eqno49}).
 We note that the integrated pion flux from pion due to
 the $s=-1$ essential residue from Eq.~(\ref{eqno53a})
 is not plotted because it is much smaller than the others.
 Also the integrated pion flux from pion due to
 the $s=s_{0},s^{*}_{0}$ simple residues in Fig.2 shows three
 bounces. This is not true. Actually, the second bounce
 corresponds to negative values, and PPs is an oscillating
 function. This is clear from the energy dependent argument
 of the sine function in Eq.~(\ref{eqno49}). Since we are
 doing a Log plot, the absolute value is taken which rectifies
 the negative part and gives the pulsating appearance.
 The overwhelming contribution of the integrated hadron flux comes
 from the PPe $s=0$ essential residue.
 The total integrated flux with $\beta=0$ is shown in Fig.3
 which presents a curvature compatible to Fig.1.
 We note that Fig.3 is the integrated flux in terms of the energy
 distribution function which could be multiplied by an arbitrary
 constant. This amounts to displace the plot along the vertical
 axis to the scale of Fig.1 for comparisons.
 To show the effects of energy dependent mean free paths,
 the total integrated flux with $\beta=0.06$ is also shown
 in Fig.3 for comparisons.

To conclude, we have pointed out in Sec.2 that the generally accepted
 diffusion equation in the Mellin transform space in terms of the
 parameter $<\eta^{s}>$, Eq.~(\ref{eqno5}) and Eq.~(\ref{eqno6a}),
 is not a mathematically rigorious representation because of the
 approximations involved in evaluating the double integral in energy.
 Since two wrongs do not make a right, the subsequent parametization
 of $<\eta^{s}>$ does not correct the mistake, and the model gets
 distorted from the start.
 Since this diffusion equation is the fundamental equation in
 evaluating fluxes of different generations where each generation
 relies on the flux of the prior flux of the parent generation,
 any misrepresentation on this basic equation would cascade
 accumulatively rendering the final results entirely meaningless.
 To overcome this primary problem, we have used the Faltung theorem
 to formulate an exact self-consistent diffusion equation for the
 flux transform, as given by Eq.~(\ref{eqno7b}) and Eq.~(\ref{eqno7c})
 in Sec.2. This Faltung formulation requires only the elasticity
 distribution, not $<\eta^{s}>$, as the primary input function
 which avoids excessive, and often conflicting, parametization.
 Energy dependence on the elasticity distribution could be included
 in this Faltung formulation.

The flux transform is solved by the method of characteristics,
 and the flux in real space is evaluated by residues, simple and
 essential. Since the essential residues come from singularities
 in the exponent which contains $\tilde{u}(s)$ and $\tilde{v}(s)$,
 this makes the choice of $u(\eta)$ and $v(\eta)$ particularly
 important. Arbitrary profile fitting could bring in more residues
 that might not be physical, such as Eq.~(\ref{eqno38}).
 Care should be exercised not to raise the power of $\eta$ in
 $\phi(E^{\prime},E)$ for fitting purposes. This would increase
 the singularties in $\tilde{v}(s)$ generating artificial fluxes.
 Although we have considered $\eta$ that is independent of the
 incident energy $E^{\prime}$, this Faltung formulation does allow
 generalizations to $\eta(E^{\prime})$ cases.
 This self-consistent Faltung formulation provides a firm and
 reliable starting point for one-dimensional cascades.

Using a method recently developed by us \cite{tsui},
 we have calculated the nucleon flux through the essential
 residue at different depths in a wide energy range initiated
 by one single nucleon. We have generalized earlier results to
 include the energy dependence in the collision mean-free path.
 Our solution is presented in the usual modified Bessel functions
 of order 1 with an energy dependent mean free path argument.
 For the pion flux, it is given by the simple and essential
 residues as well. A comparison of the integrated hadron flux
 with the hadronic spectra measured at Mt. Chacaltaya for
 four halo events shows good agreements even with $\beta=0$.
 Our hadron flux presents a curvature on the Log-Log plot
 just like the observations, althought we have used a simple
 uniform elasticity distribution.
 The fundamental reasons of this good agreement stem from three aspects.
 The first is the exact representation of the diffusion equation
 through Faltung formulation with elasticity distribution as the
 basic input function. This exact formulation eliminates unnecessary
 free parameters that only serve to contaminate and distort models
 such as $<\eta^{s}>$ which is forced upon by a bad derivation.
 The second is the use of essential residues to evaluate the flux
 which has been overlooked by earlier investigators.
 The third is the use of elasticity distribution $u(\eta)$ and
 consequently the self-consistent $v(\eta)$ to evaluate the flux.
 Because of these three aspects, our model with no free parameters
 gives results compatible to observations using only a simple
 uniform elasticity distribution.


\newpage

\begin{figure}[th]
\vspace*{-0.0cm}
\includegraphics[clip,width=0.9
\textwidth,height=0.9\textheight,angle=0.] {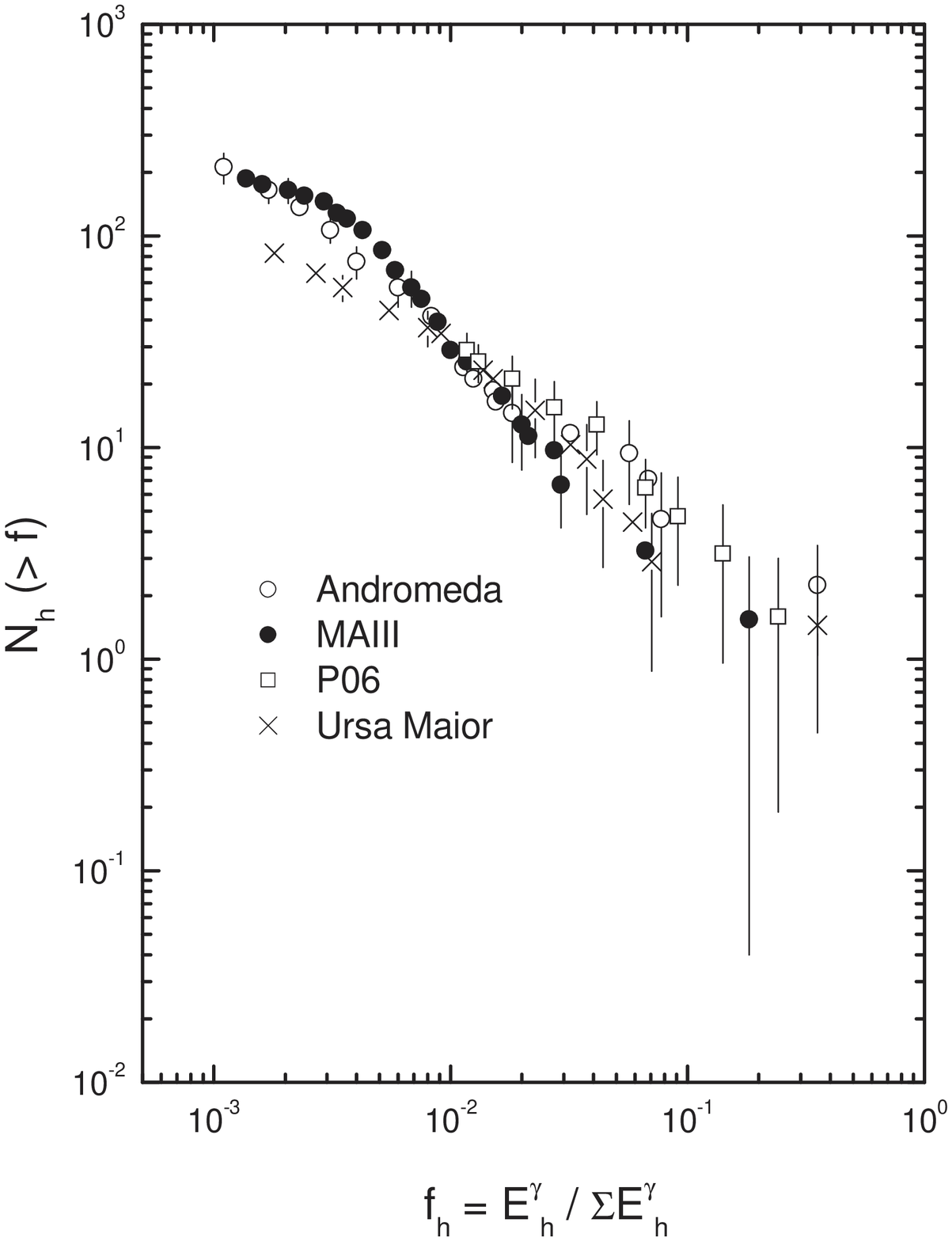}
\vspace*{-1.0cm}
\caption{The experimental data of four halo events at Mt. Chacaltaya.}
\end{figure}

\newpage

\begin{figure}[th]
\vspace*{-0.0cm}
\includegraphics[clip,width=0.9
\textwidth,height=0.9\textheight,angle=0.] {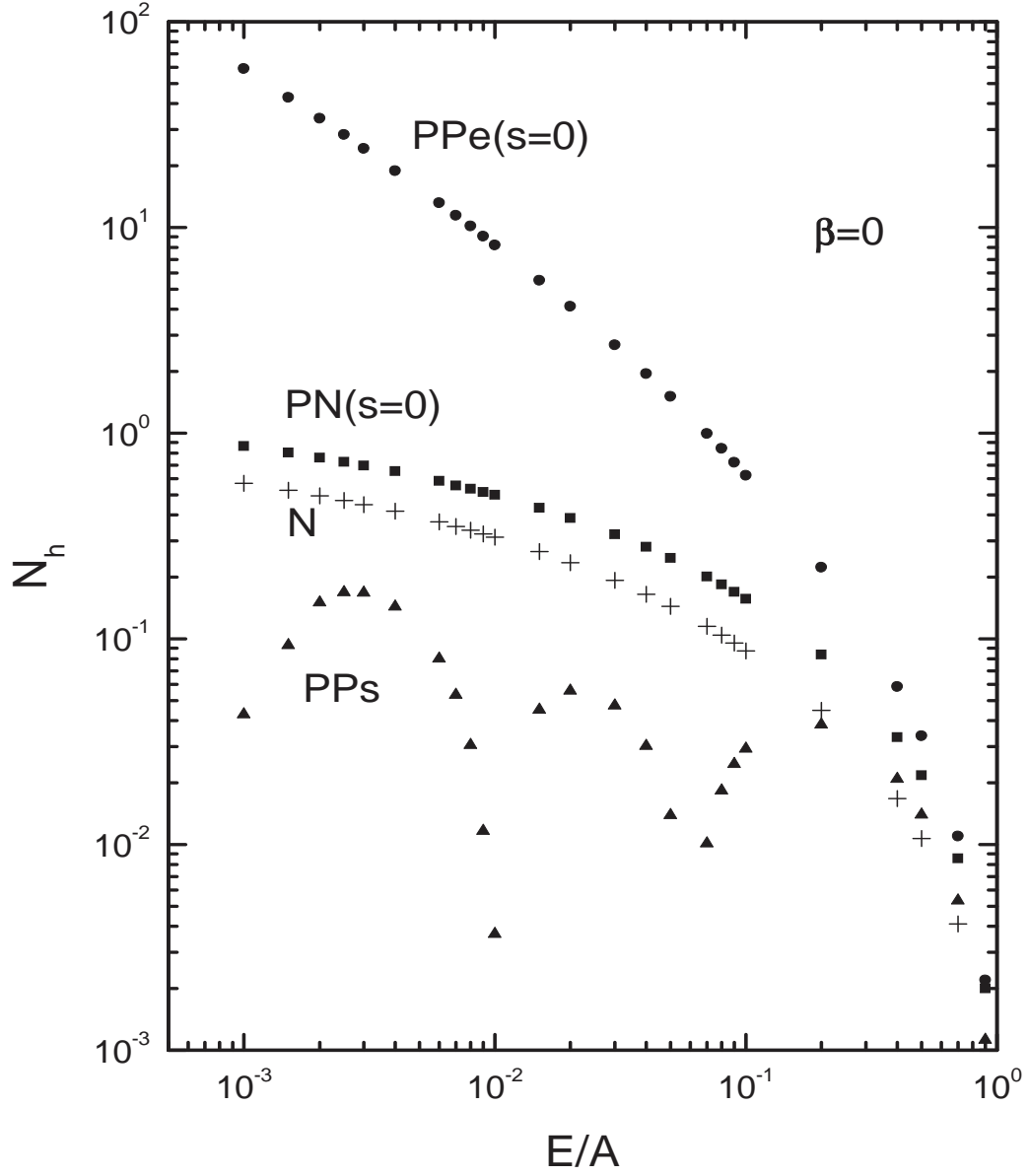}
\vspace*{-1.0cm}
\caption{The $\beta=0$ integrated flux of the nucleon (N),
 the pion from nucleon of the essential residue $s=0$ (PN),
 the pion from pion of the simple residue (PPs),
 the pion from pion of the essential residue $s=0$ (PPe).}
\end{figure}

\newpage

\begin{figure}[th]
\vspace*{-0.0cm}
\includegraphics[clip,width=0.9
\textwidth,height=0.9\textheight,angle=0.] {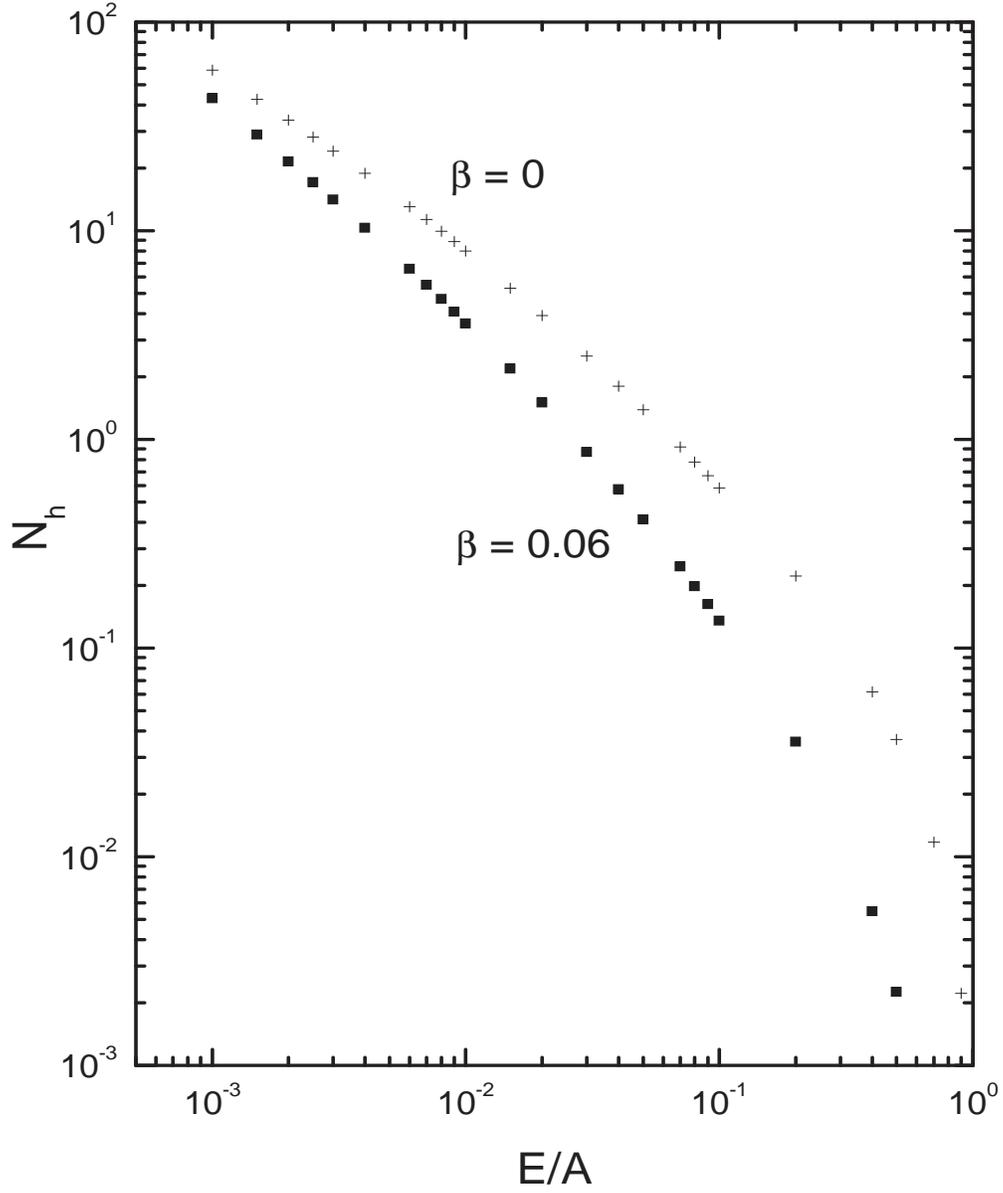}
\vspace*{-1.0cm}
\caption{The $\beta=0$ and $\beta=0.06$ total integrated pion flux.}
\end{figure}

\end{document}